\documentclass[useAMS,usenatbib]{mn2e}
\usepackage[colorlinks]{hyperref}
\usepackage{amssymb}
\usepackage{amsmath}
\usepackage{txfonts}
\usepackage{graphicx}
\usepackage{subcaption} 
\def\gr{$\gamma$-ray}

\usepackage{epsfig}
\usepackage{xspace}
\usepackage{url}
\usepackage[colorinlistoftodos]{todonotes}

\usepackage{natbib}
\def\flat{\textit{Fermi/LAT}\xspace}
\def\lsi{LS I +61$^\circ$303\xspace}

\def\psrb {PSR~B1259-63\xspace}
\def\psrj {PSR~J2032+4127\xspace}

\begin{document}

\title{Energy-dependent periodicities of LS I +61$^\circ$ 303 in the GeV band}
\author[Chernyakova et al]{M.~Chernyakova$^{1,2}$, D.~Malyshev$^3$, A.~Neronov$^{4,5}$, D.~Savchenko$^{4,6,7}$\\
$^{1}$  School of Physical Sciences and Centre for Astrophysics \& Relativity, Dublin City University, D09 W6Y4 Glasnevin, Ireland\\
$^2$ Dublin Institute for Advanced Studies, 31 Fitzwilliam Place, D02 XF86 Dublin 2,  Ireland\\
$^3$ Institut f\"ur Astronomie und Astrophysik, Universit\"at T\"ubingen, Sand 1, D 72076 T\"ubingen, Germany\\
$^{4}$Universit\'e de Paris, CNRS, Astroparticule et Cosmologie, 
F-75006 Paris, France\\
$^{5}$Laboratory of Astrophysics, \'Ecole Polytechnique F\'ed\'erale de Lausanne, CH-1015 Lausanne, Switzerland\\
$^{6}$Bogolyubov Institute for Theoretical Physics of the NAS of Ukraine, 03143 Kyiv, Ukraine\\
$^{7}$Kyiv Academic University, 03142 Kyiv, Ukraine}

\maketitle
\begin{abstract}
\lsi is a rare representative of the gamma-ray binaries with a compact object known to be a pulsar. We report on the periodicity and spectral analysis of this source performed with more than 14 years of \flat data. The periodicity of \lsi is strongly energy dependent. Two periods $P_1 = 26.932\pm  0.004 (stat)\pm 0.008 (syst)$ and $P_2 = 26.485 \pm 0.004 (stat)\pm 0.007 (syst)$ are detected only at $E>1$~GeV and at $E<0.3$~GeV correspondingly. Within $1\sigma$ (stat+syst) the periods are consistent with orbital ($P_2$) and beat orbital/superorbital ($P_1$) periods. We present the orbital light curves of the system in several energy bands and the results of the spectral analysis. We discuss the possible origin of the change in the variability pattern between 0.1 and 1 GeV energy.
\end{abstract}
\begin{keywords}
(stars:) binaries:general -- (stars:) pulsars: individual: LS~I~+61$^\circ$303 -- gamma-rays: stars    
\end{keywords}

\section{Introduction}
Gamma-ray-loud binary systems (GRLB) are X-ray binaries which emit very-high energy (VHE) $\gamma$ rays. While about a thousand of X-ray binaries are known, only about a dozen of systems have been detected as persistent or regularly variable GeV-TeV emitters~\citep[see e.g.][]{dubus13, we_review}. 


 The power engine (accretion or rotation powered), the physical conditions allowing the acceleration of charged particles to the very high energies (and consequent very high energy photon emission) and even the nature of the compact object (CO) are not well established for almost all GRLBs. E.g. the absence of the pulsed radio emission from some systems can point to the black hole nature of the compact object. On the other hand the detection of the pulsed radio emission can be complicated by a strong absorption of such an emission in the dense layers of the stellar decretion disk. 
 

Among all $\gamma$-ray binaries the type of the compact object was firmly established to be a pulsar only for three systems. Until recently, the CO was identified (through detection of the pulsed radio emission) as a pulsar only in \psrb\ and \psrj~\citep{johnston92}. In 2022  Five-hundred-meter Aperture Spherical radio Telescope (FAST) radio observations allowed the detection of the pulsed radio emission from \lsi~\citep{lsi_pulsar}, increasing the number of pulsar-hosting systems to three.


\lsi consists of a Be star and a pulsar on the eccentric orbit. Two decade-long radio observations of \lsi demonstrated that the emission is modulated on timescales of $P_o\sim 26.5$~d and $P_{so}\sim 1667$~d~\citep{gregory02}, referred hereafter as orbital and superorbital periods. Similar periods have been detected in optical~\citep{zamanov13,fortuny15}, X-ray~\citep{we_lsi_xray,li14} and gamma-ray bands~\citep{lsifermi13,jaron14}. In radio to X-ray bands, the orbital light curve is characterized by a single peak with a wavelength-dependent position and drifting on the superorbital time scale. With the change of superorbital phase the peak demonstrates rapid transition to another orbital phase~\citep{we_lsi_xray}. At the same time, the behaviour of the system in the GeV band seems to differ significantly, with the structure of the orbital light curve changing from a regular with a single peak at certain superorbital phases to erratic at others~\citep{we_lsi_xray,saha16,xing17}. 

In this paper, we reconsider public GeV-band observations of \lsi aiming for detailed studies of the variability of this system on orbital and superorbital time scales. We find that the periodicity properties of the system are strongly energy-dependent in the energy range accessible to the \flat instrument.

\begin{figure*}
\includegraphics[width=\columnwidth]{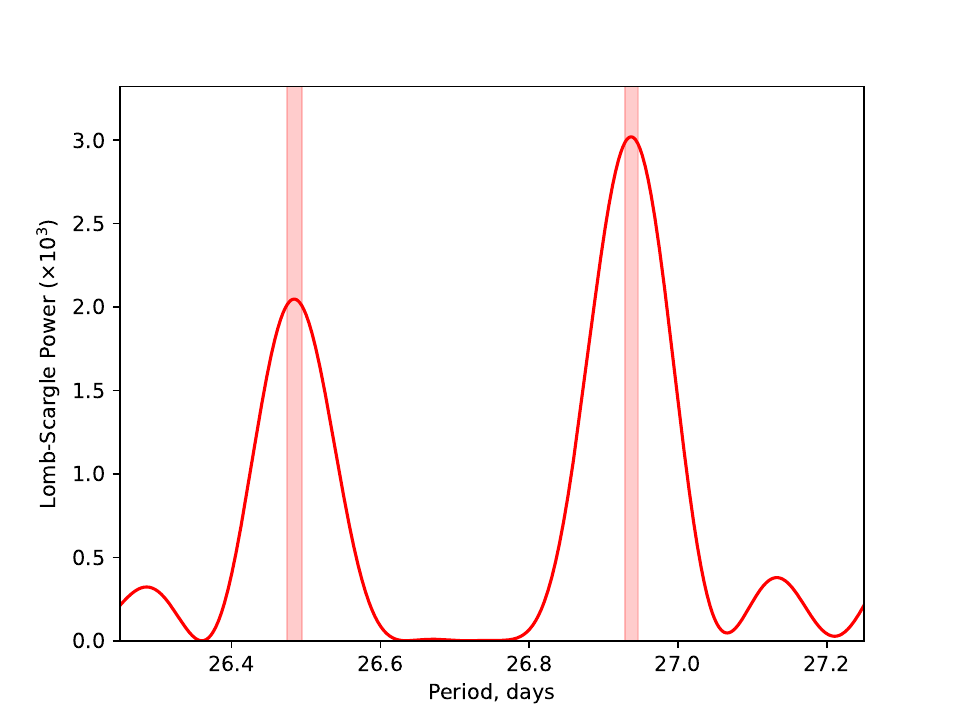}
\includegraphics[width=\columnwidth]{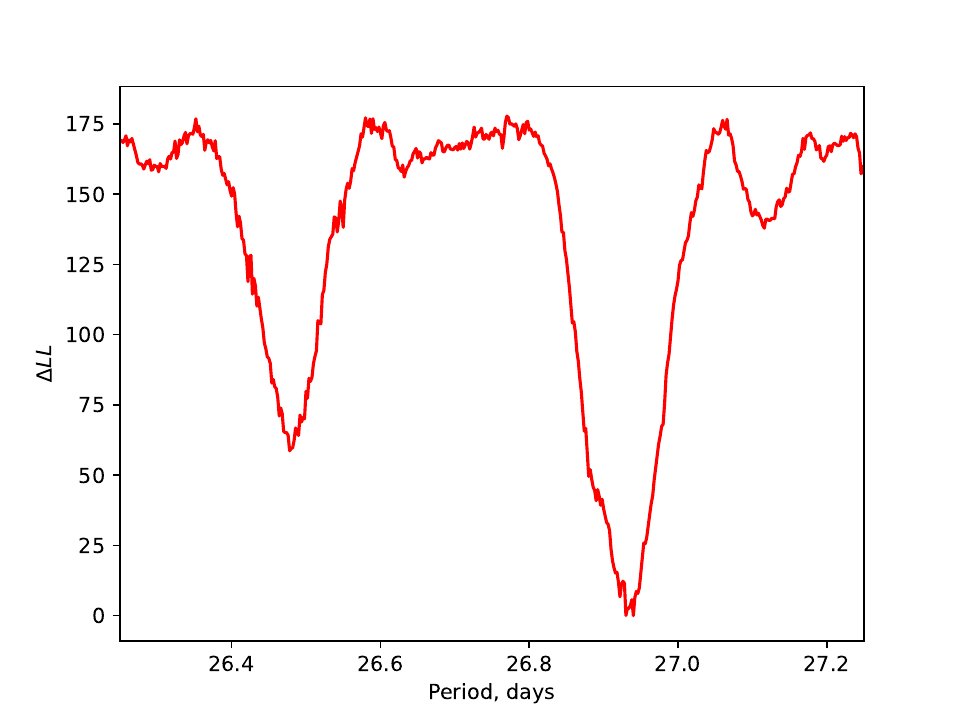}
\caption{\textit{Left}: Lomb-Scargle period search with 0.3-300~GeV \flat data. Shaded intervals illustrate $2\sigma$ confidence ranges around the positions of $P_1$ and $P_2$ periods. \textit{Right}: Log-like profile for the self-similar searches of periodicity in Fermi/LAT data. Orbital (left) and ``superorbital'' (right)  periods correspond to the minima in LL profile.
 }
\label{fig:period_searches}
\end{figure*}

The paper is organised as follows: in Sec.~\ref{se:lat_data_analysis} we discuss the \flat data and methods used for its analysis; in Sec.~\ref{sec:results} we present the obtained results and discuss the possible origin of the observed energy-dependent periodicity. In Sec.~\ref{sec:conclusions} we shortly summarize the obtained results and their possible interpretation.

Where applicable in what below, we adopt the following parameters -- the orbital and superorbital periods $P_{orb} = 26.496 \pm 
0.0028$ d and $P_{sorb} = 1667$~d~\citep{gregory02}. The values for the eccentricity of $e = 0.537 \pm 0.034$ and the phase of the periastron of $\phi = 0.275 \pm 0.010$ are adopted from~\citet{aragona09}, see however~\citet{kravtsov20}. Historically the phase of $\phi = 0$  corresponds to Julian Date
(JD) 2,443,366.775~\citep{gregory02}.

\section{\flat Data and Data Analysis}
\label{se:lat_data_analysis}
The results described below are based on the analysis of more than 14 years of the \flat data (Aug. 4th, 2008 -- Oct. 26th, 2022) with the latest available \texttt{Fermitools v.2.0.8} software. 
The analysis was carried out using the latest Pass 8 reprocessed data (P8R3, \citet{atwood2013pass}) for the CLEAN event class taken at the region centered at \lsi coordinates. Further details specific to the performed analysis are summarized in corresponding subsections.

\subsection{\flat data: aperture photometry analysis}
\subsubsection{Periodicity searches}
Aiming in periodicity studies in \lsi data we build the light curves of \lsi with the standard aperture photometry analysis\footnote{See e.g. \href{https://fermi.gsfc.nasa.gov/ssc/data/analysis/scitools/aperture_photometry.html}{\flat aperture photometry analysis manual}} in several energy intervals.
In each of the considered energy intervals (0.1--0.3\,GeV; 0.3--1\,GeV; 1-10\,GeV; 0.3-300\,GeV) we selected the photons with the corresponding energies, detected within $1^\circ$-radius around \lsi. We binned the selected photons into lightcruves with 30~min long time bins and calculated the exposure for each time bin with the help of \texttt{gtbin} and  \texttt{gtexposure} routines. Please note, that for this type of the analysis we explicitly built background not-subtracted lightcuve.

\begin{figure*}
\includegraphics[width=\columnwidth]{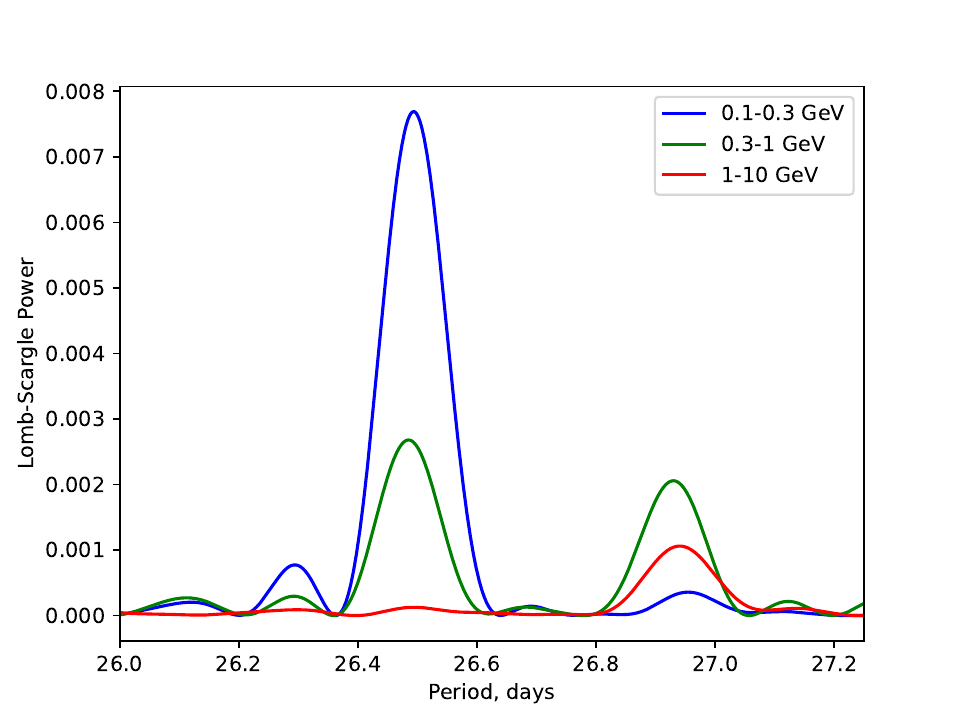}
\includegraphics[width=\columnwidth]{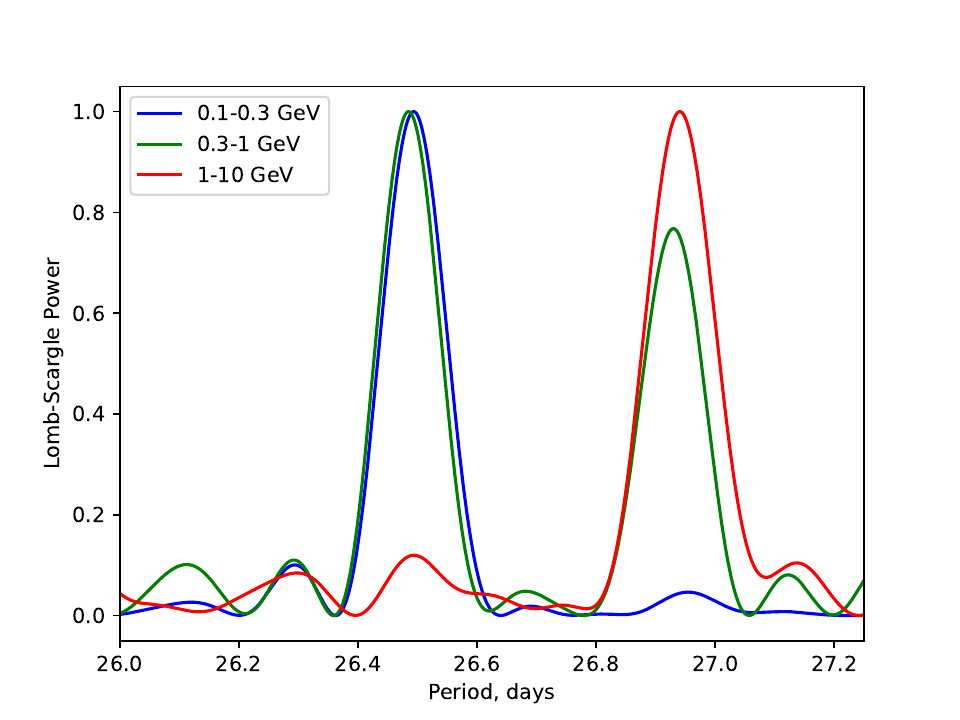}
\caption{left panel: Results of Lomb-Scargle analysis performed on \flat light curve (aperture analysis,$1^\circ$-radius, 30~mins time bins) at several energy bands. Right panel: Same, normalized to 1 at maxima.}
 \label{fig:LS_3energies}
\end{figure*}

\subsubsection{Lomb-Scargle analysis}
We performed the Lomb-Scargle analysis of the obtained light curves using the implementation  provided within \texttt{astropy (v.5.1)} python module. The time bins with zero \flat exposure were explicitly removed during the analysis. The periodograms were built for 2000 trial periods linearly distributed between $26-28$~days. 

A zoom of the periodogram in the  energy range (0.3--300\,GeV) to $26.25-27.25$~d period range is shown in Fig.~\ref{fig:period_searches} (left). 
Two periods $P_1 = 26.937$~d and $P_2=26.4845$~d corresponding to local maxima at the periodogram are clearly visible in the Figure. 

In order to estimate the uncertainty of these periods we performed the Lomb-Scargle analysis of $10^3$ randomly-generated datasets. Each random dataset was generated according to a Poisson distribution around the real dataset. In each of the generated datasets, we determined the positions of local maxima close to $P_1$ and $P_2$ positions. The distribution of maxima allowed us to estimate ($1\sigma$) uncertainty for these periods as $P_1 = 26.932\pm  0.0042$~d and $P_2 = 26.4845 \pm 0.0046$~d. We note that $P_2$ period is consistent with $P_{orb}$ at $\sim 2\sigma$ level. The period $P_1$ is at $\sim 2\sigma$ level consistent with 
the orbital-superorbital beat-period ($P_{beat}=P_{orb}P_{sorb}/(P_{sorb}-P_{orb})\simeq 26.924$~d).

The shaded regions around $P_1$ and $P_2$ positions in the left panel of Fig.~\ref{fig:period_searches} correspond to $2\sigma$ statistical uncertainty regions derived from the random datasets as described above.

The Lomb-Scargle periodograms built in narrower energy intervals (0.1--0.3\,GeV; 0.3--1\,GeV; 1-10\,GeV) are shown in Fig.~\ref{fig:LS_3energies}. These periodograms demonstrate clear energy dependence of the Lomb-Scargle power in the peaks corresponding to $P_1$ and $P_2$ periods. While at the lowest (0.1--0.3\,GeV) energies the $P_2$ period dominates the periodogram, at highest energies (1--10\,GeV) the periodogram is dominated by $P_1$. At intermediate energies (0.3--1\,GeV) both periods are clearly seen.

\subsubsection{Self-Similar Log-Likelihood analysis}
To cross-check the results of Lomb-Scargle analysis discussed in the previous subsection we additionally performed the self-similar log-likelihood analysis. A similar analysis was shown to be effective for the blind search of the periodicity in the \flat data of gamma-ray binary 1FGL J1018.6-5856~\citep{soelen22}. 

For the analysis, we first defined  a range of the test periods (1000 periods linearly distributed between 26~d and 28~d). We convolved the \flat light curve with each of the test periods defining corresponding ``test orbital light curves'' assuming 20 linearly distributed phase-bins per orbit. Based on each test orbital light curve we defined the predicted (``model'') number of counts in each time bin of the original light curve. In the next step, we calculated the log-likelihood to observe the detected number of photons in each time bin $N_i$ given the model number of photons $m_i$:
\begin{align}
 & \log LL = \sum\limits_i \log P_i(\geq N_i\mid m_i)   
\end{align}
Here $P_i(\geq N_i\mid m_i)$ stands for the Poisson probability to observe $\geq N_i$ photons if the model predicts $m_i$ photons in the time bin $i$. Figure~\ref{fig:period_searches} (right panel) shows the $\Delta LL = -2\cdot(\log LL - \min(\log LL))$ profile for the considered test periods. The quantity $\Delta LL$ follows the $\chi^2$ distribution with 1~d.o.f.~\citep{wilks38} and can be used thus for the estimation of the periods present in the data and uncertainties on these periods.

The self-similar log-likelihood analysis resulted in the detection of two periods similar to the Lomb-Scargle approach. The corresponding periods are $P_1=26.940 \pm 0.006$~d and $P_2=26.478 \pm 0.003$~d. 

We use this independent from Lomb-Scargle analysis to estimate the level of systematic uncertainty of the performed analysis. Namely, we require the best-fit $P_1$ and $P_2$ values obtained from Lomb-Scargle and self-similar log-likelihood analysis to be consistent within $1\sigma$ systematic uncertainty. This results in the following estimations of the periods: $P_1 = 26.932\pm  0.004 (stat)\pm 0.008 (syst)$~d and $P_2 = 26.485 \pm 0.004 (stat)\pm 0.007 (syst)$~d.

\begin{figure*}
\includegraphics[width=0.65\columnwidth]{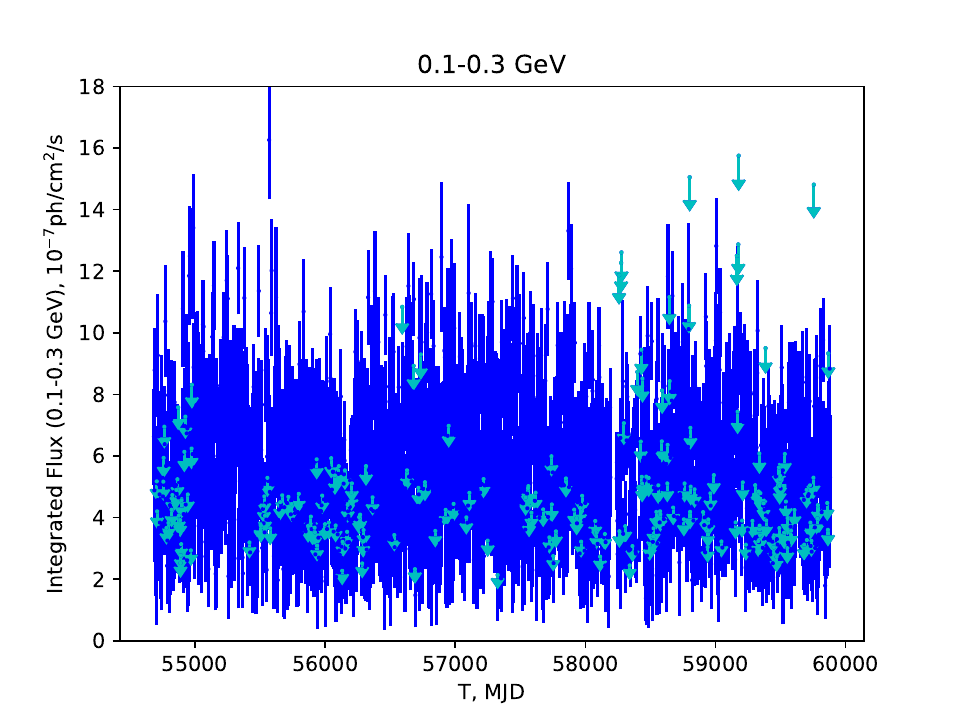}    
\includegraphics[width=0.65\columnwidth]{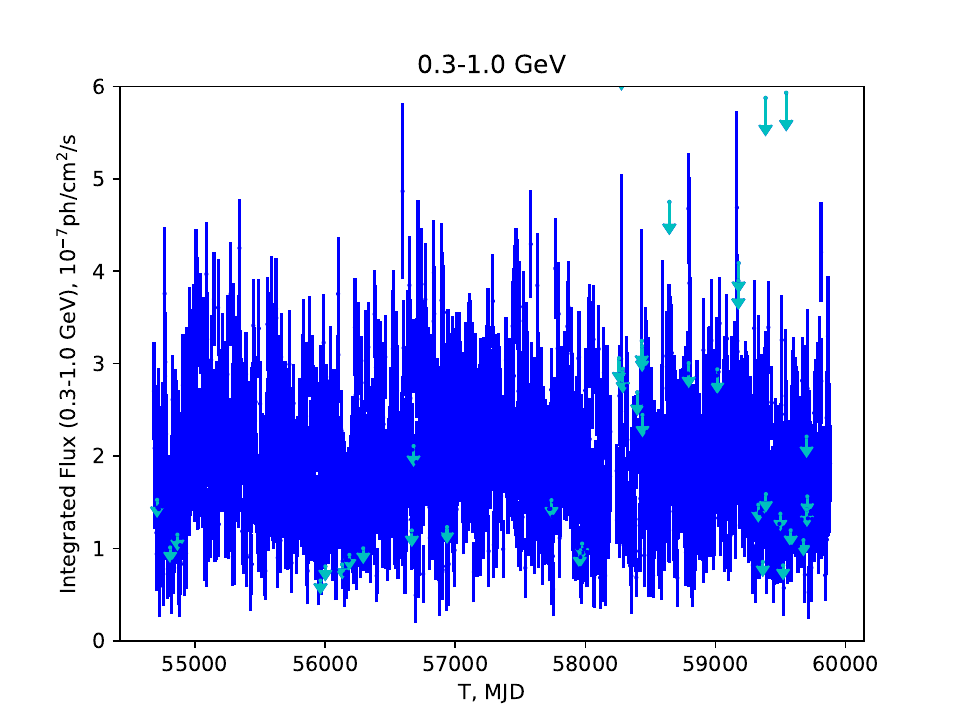}
\includegraphics[width=0.65\columnwidth]{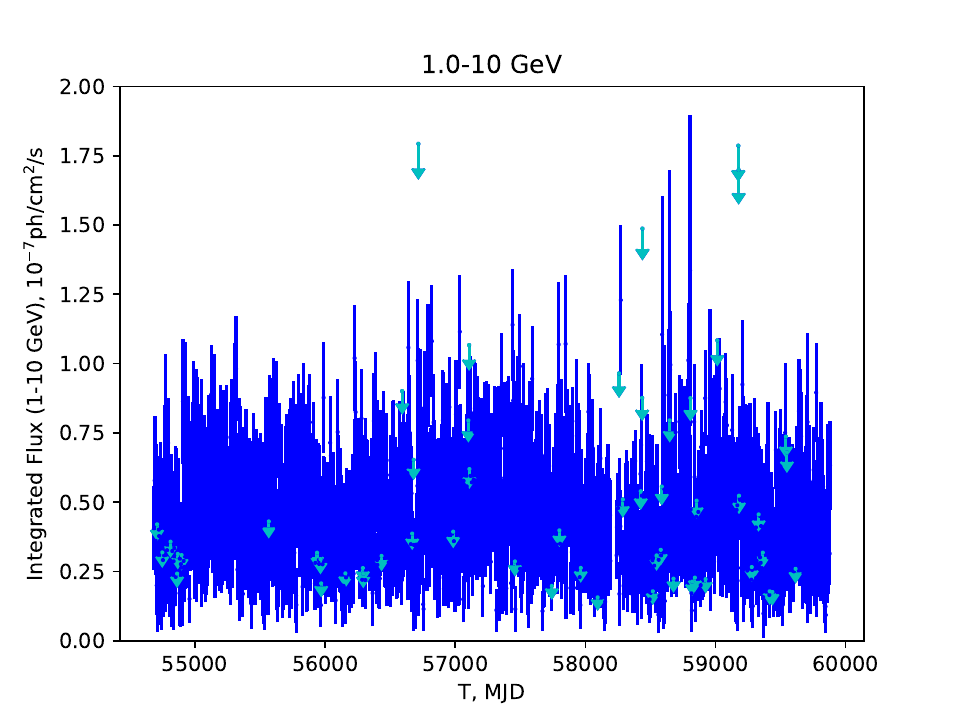}

\caption{The likelihood analysis (background subtracted) lightcurves of \lsi in 0.1-0.3 GeV (left panel), 0.3-1.0 GeV (middle) and 1.0-10~GeV (right) energy bands. All time binning for all lightcurves is $2.6496$~d. (0.1 orbital period). The upper limits are shown with cyan and correspond to the 95\% confidence range.}
\label{fig:fermilc}
\end{figure*}

\begin{figure*}
\includegraphics[width=\columnwidth]{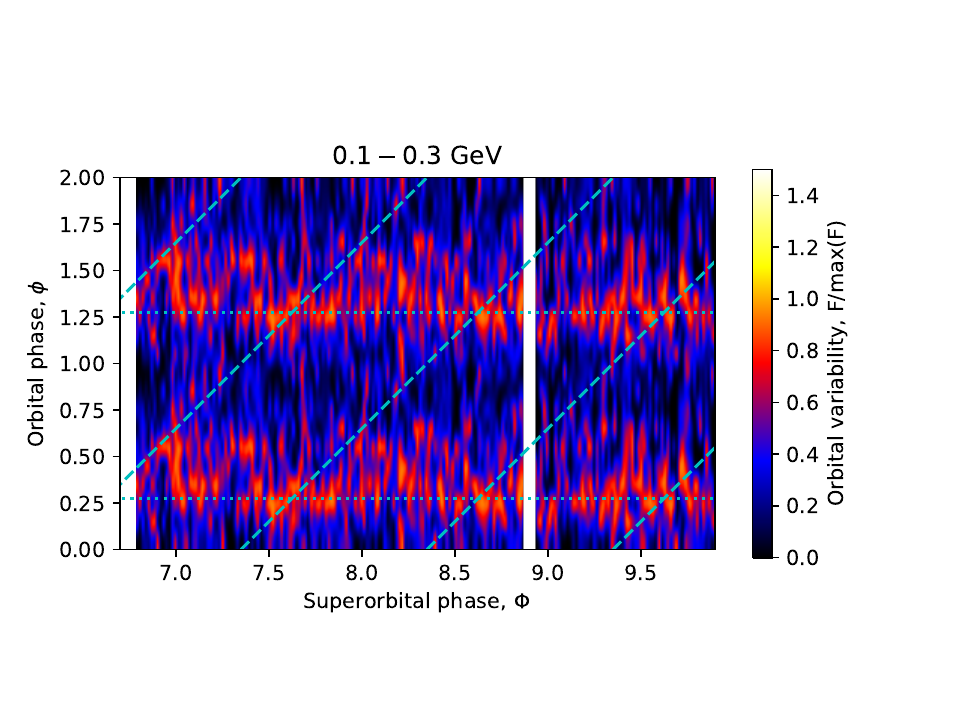}    
\includegraphics[width=\columnwidth]{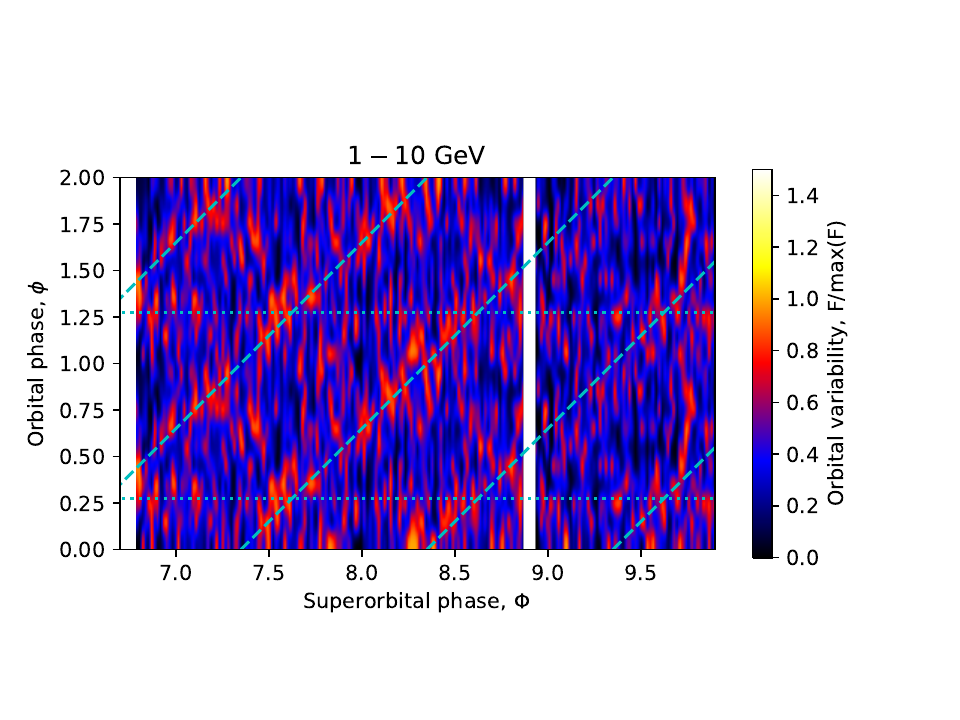}
\caption{The orbital variability of the \lsi flux as a function of orbital and superorbital phases in 0.1-0.3~GeV and 1-10~GeV energy bands. The dotted horizontal cyan line shows the orbital phase of the periastron ($\phi=0.275$). Dashed diagonal cyan lines illustrate the position of the GeV flux peak connected to $P_{beat}$ period. A solid green line illustrates the position of the X-ray maximum at corresponding orbital/superorbital phases.}
\label{fig:2dlc}
\end{figure*}

\subsection{\flat data: likelihood analysis}
\label{sec:likelihood_analysis}
To study the details of the variability of flux and spectral characteristics of the source on the orbital and beat period time scales, we additionally performed the standard binned likelihood analysis of \flat data. Contrary to the aperture photometry analysis such analysis relies on the fitting of the available data to the spatial/spectral model of the analysed region and allows to account for possible flux variations of the nearby sources. 

For the binned likelihood analysis~\citep{mattox96} we consider the photons (CLEAN class, P8R3 IRFs, $zmax=90^\circ$) within a circular region of $12^{\circ}$-radius centered at \lsi position. The spatial/spectral model of the region included the standard galactic and isotropic diffuse emission components as well as all known gamma-ray sources within $17^{\circ}$ of the ROI centre from the 4FGL catalogue~\citep{4FGL}. Namely, the positions and spectral models for each catalogue source were selected according to the ones provided in the catalogue. During the fitting of the model to the data in each of the time bins only the normalisations of spectral models were left free for all sources, while all other spectral parameters were fixed to their catalogue values. We additionally fixed all spectral parameters (including normalisations) of all sources in a ring $12^\circ-17^\circ$ from \lsi position. We note, that within this analysis we have explicitly built the background subtracted lightcurve. The background model in this case included contributions for the nearby 4FGL catalogue sources as well as contributions from the galactic and extragalactic backgrounds.

\subsubsection{Orbital light curves}
\begin{figure*}
\includegraphics[width=\columnwidth]{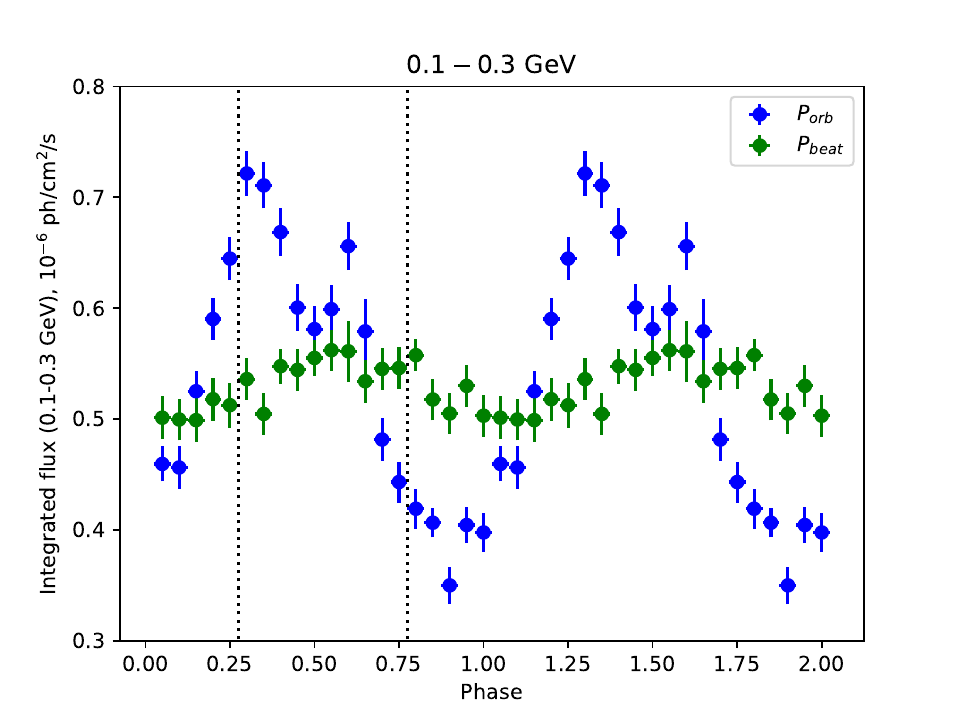}
\includegraphics[width=\columnwidth]{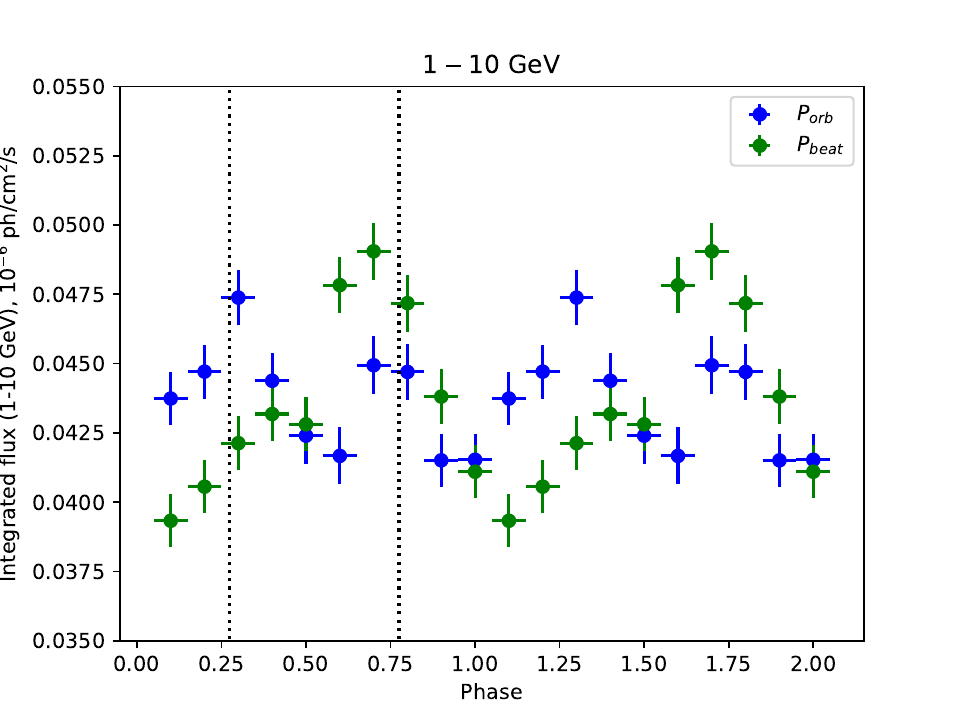}
\caption{light curves of \lsi as seen by \flat (binned likelihood analysis) convolved with orbital and beat-periods in different energy bands. Vertical lines show periastron ($\phi=0.275$) and apastron positions.}
\label{fig:orbital_lc_multien}
\end{figure*}

In order to build the folded orbital light curves we split the data into energy bins (0.1--0.3\,GeV and 1--10\,GeV) and time bins (according to orbital phase with respect to $P_{orb}$ or $P_{beat}$) and the analysis was performed in each of such bins. Zero phases in both cases were selected to be $T_0 = MJD\,43366.275$ as in previous studies of orbital and superorbital periodicity of \lsi. The orbital light curves produced in this way are shown in Fig.~\ref{fig:orbital_lc_multien}. The left panel shows the folded light curves for the energy range 0.1--0.3\,GeV, with the period $P_{orb}$ (blue points) and $P_{beat}$ (green points). The right panel shows the  results for the $1-10$\,GeV energy range. Vertical dashed lines correspond to the periastron ($\phi=0.275$) and the apastron phases. 

In addition to the folded orbital light curves we performed studies of the long-term variations of the orbital light curve (with respect to $P_{orb}$) with the $P_{sorb}$. For this, we have performed the  binned likelihood analysis in the specified energy intervals and time bins as short as 2.6496~d. (i.e. 0.1 orbital phase duration) aiming to determine \lsi flux in each of the specified time bins. The obtained ligtcurves in 0.1-0.3~GeV, 0.3-1.0~GeV and 1.0-10~GeV are shown in Fig.~\ref{fig:fermilc} with blue color. The cyan points correspond to the 95\% c.r. flux upper limits (calculated with \texttt{IntegralUpperLimit} module included into fermi tools).

Fig.~\ref{fig:2dlc} shows the fractional variability of the \lsi flux (i.e. flux normalized to 1 at each orbit by the maximal flux observed at that orbit) in 0.1-0.3~GeV and 1-10~GeV energy intervals as a function of orbital and superorbital phases. In case of no significant detection of the source in a time bin, we explicitly set the flux in this time bin to 0.

The white gaps correspond to the period in March-April 2018. During this period \flat was in the safe hold mode due to issues with the solar array and did not take scientific data~\footnote{See e.g. \href{https://www.nasa.gov/feature/goddard/2018/fermi-status-update}{\flat status report}}.
Dotted horizontal cyan lines correspond to the phase of the periastron ($\phi=0.275$). Dashed diagonal cyan lines illustrate the position of flux maximum seen above 1~GeV energies. This maximum is drifting with respect to $P_{orb}$ and can be connected to the beat period $P_{beat}$. 

Overall, we identify  several distinct orbital/superorbital phase periods. These are:
\begin{itemize}
    \item periastron maximum ($\phi=0.275\pm 0.1$); observed in 0.1-0.3~GeV range;
    \item ``2nd peak'' or beat-period maximum ($frac(\Phi)=(frac(\phi)-0.35)\pm 0.1 $)  along dashed diagonal lines,  seen above 1~GeV;
    \item ``minima'': periods of low GeV emission at $E<0.3$~GeV ($frac(\Phi)>0.4$ AND $frac(\phi)>0.75$).
\end{itemize}
Here $frac$ stands for the fractional part of the orbital($\phi$) or superobital($\Phi$) phase.

For each of these time intervals,  we  performed the binned likelihood analysis for a set of energy bins. The best-fit fluxes for the corresponding energy/time bins as a function of energy are shown in Fig.~\ref{fig:lat_spectra}. The upper limits on the flux are shown for the energy bins where \lsi is not detected with at least $2\sigma$ significance. The upper limits correspond to 95\% false-chance probability and are calculated with the help of \texttt{IntegralUpperLimit} python module, provided within \texttt{Fermitools}.

\section{Results and Discussion}
\label{sec:results}

Our analysis of more than 14~yr of \flat data on \lsi has revealed the presence of two close-by periods  $P_1 = 26.932\pm  0.004 (stat)\pm 0.008 (syst)$~d and $P_2 = 26.485 \pm 0.004 (stat)\pm 0.007 (syst)$~d, see Fig.~\ref{fig:period_searches}. Within $1\sigma$ (stat+syst) uncertainties these periods coincide with the orbital period ($P_2\approx P_{orb}=26.496$~d) and the orbital-superorbital beat-period ($P_1\approx P_{beat}=P_{orb}P_{sorb}/(P_{sorb}-P_{orb})\simeq 26.924$~d).

The periodicity of the signal is strongly energy dependent. The low-energy ($0.1-0.3$~GeV) light curve is strongly modulated with an orbital period and peaks at close to periastron phases. At higher, $1-10$~GeV energies, the modulation with the orbital period is strongly suppressed and only orbital/superorbital beat period is detected, see Fig.~\ref{fig:LS_3energies}. The light curves folded with $P_{orb}$ and $P_{beat}$ periods are shown in Fig.~\ref{fig:orbital_lc_multien}.
The lower energy band orbit-folded light curve has a clear peak at the periastron ($\phi=0.275$) and a secondary peak at the orbital phase $\phi\sim 0.6$. 

The $1-10$~GeV orbit-folded light curve still possibly features the peak at the periastron, even though the amplitude of the orbital variability is strongly decreased. Instead, a pronounced peak is found in the beat-period folded light curve. 

The energy-dependent  variability pattern is also clearly seen in Fig.~\ref{fig:2dlc} showing the level of the orbital variability as a function of orbital and superorbital phases. In this figure, the orbital phase of the periastron peak is shown with the dotted horizontal line and the beat period seen at higher energies corresponds to the diagonal dashed cyan lines. It corresponds to a gradual drift of the phase of the maximum of the $E>1$~GeV light curve throughout the super-orbital cycle. 

\begin{figure}
\includegraphics[width=\columnwidth]{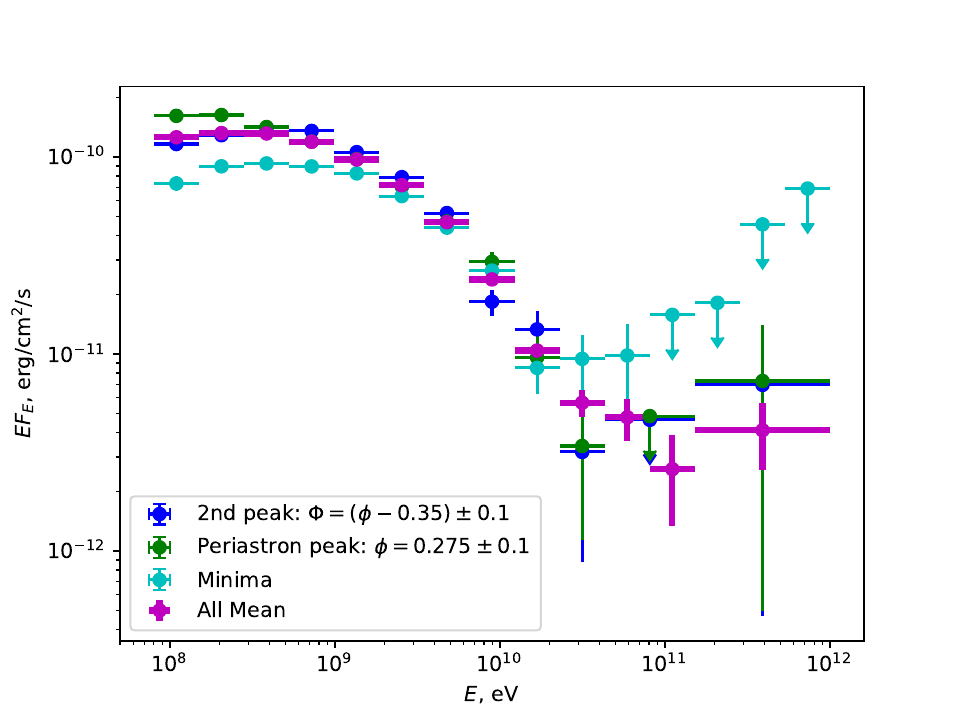}
\caption{\flat spectra of \lsi at different orbital periods}
\label{fig:lat_spectra}
\end{figure}

The spectra extracted at phases around the periastron peak ($\phi=0.275\pm 0.1$), beat-period maximum (``2nd peak'', $frac(\Phi)=(frac(\phi)-0.35)\pm 0.1$), minimal low-energy flux ($frac(\Phi)>0.4$ AND $frac(\phi)>0.75$) and the all-data averaged spectra are shown in Fig.~\ref{fig:lat_spectra}. One can see that the source is most variable in the energy range below 1~GeV.  The peak flux energy changes from about 0.1~GeV at the periastron  to $\sim 0.4$~GeV for the beat-period maximum and minimal low-energy flux periods. Surprisingly, the flux in the energy range above the peak is more stable than below the peak. 

The presence of the peak in $0.1-0.3$~GeV orbital light curve at close to periastron orbital phases can be explained in a straightforward manner. The $0.1$~GeV emission could be produced via synchrotron or the IC mechanisms that naturally result in the enhanced level of $\gamma$-ray emission close to periastron due to the increased magnetic and/or soft photon fields densities at these orbital phases.

The drift of the maximum of the $E>1$~GeV light curve may possibly be explained as being due to the precession of the system components. There are multiple rotating components in the system. The fastest rotator is the pulsar that spins with the period $P_p\approx 270$~ms. The Be star spins much slower, with a period $P_*$ that should be close to the period of Keplerian orbits at the surface of the star, 

$$P_*\simeq 2\pi R_*^{3/2}/(GM)^{1/2}\simeq 0.7 (R_*/10R_\odot)^{3/2}(M/30M_\odot)^{-1/2}\,d$$

The star is surrounded by the deccretion disk that spins with nearly Keplerian velocity. This Keplerian velocity decreases with the distance from the star and is lowest at the truncation radius of the disk, which is close to the size of the binary orbit. The period of rotation at the disk truncation radius is 
$$P_{disk}\simeq 2\pi R_{disk}^{3/2}/(GM)^{1/2}\simeq 36.45(R_{disk}/10^{13}cm)^{3/2}(M/30M_\odot)^{-1/2}\,d$$ 
Finally, the binary orbit has the period $P_{orb}\simeq 26.5$~d. It can be close to the period of the disk if the truncation radius of the disk is close to the binary separation between the pulsar and the star. 
The periods $P_1$ and $P_2$ may be potentially associated to $P_*, P_{disk}, P_{orb}$ or a certain combination of these periods.  

\citet{2013A&A...554A.105M} have discussed the possibility that the  periods  $P_1$ and $P_2$ correspond to the orbital period and precession period, presumably, of a jet emitted by the black hole. Evidence for the presence of a pulsar in the system~\citep{lsi_pulsar} disfavors the hypothesis of a precessing black hole jet. Nevertheless, precession can well be relevant also for the system where the compact object is a pulsar. 

One possibility is precession due to a misalignment of the orbital plane and the equatorial plane of the Be star, which is the middle plane of the deccretion disk of the Be star. In this case, the gravitational pull of the pulsar produces a torque on the disk, which forces it to precess around the axis perpendicular to the orbital plane. The effect is similar to the precession of the rotation axis of the Earth due to the gravitational pull of the Sun. If the disk spins with the angular velocity $\vec \omega$, with the spin axis (the rotation axis of the Be star) inclined at an angle $\theta$ with respect to the orbital plane of the binary,  the precession angular frequency is \citep{2012Precession}
\begin{equation}
    \Omega_{pr}=\frac{3\epsilon\Omega_{orb}^2}{2\omega}\cos\theta
\end{equation}
where $\epsilon=(I_{||}-I_{\bot})/I_{||}$ is the ellipticity of the disk that depends on its momenta of inertia parallel and perpendicular to the rotation axis. The decretion disk is rotating with a frequency close to the frequency of rotation along Keplerian orbits at the disk truncation radius, which is close to the binary separation distance. Thus, $\omega$ is close to $\Omega_{orb}$ and $\Omega_{pr}$ can be close to both $\omega$ and $\Omega_{orb}$ for certain $\theta$ if $\cos\theta\simeq 2/(3\epsilon)$. In this case, the disk precession is almost synchronized with the pulsar rotation and a small mismatch between the disk precession and binary orbit periods leads to a gradual change of orientation of the disk with respect to the orbital plane on a long superorbital time scale. This slow change of mutual orientation of the decretion disk and pulsar may lead to the slow superorbital variability with the period $P_{so}\simeq P_{orb}^2/(P_{pr}-P_{orb})$. 

An alternative explanation may be the periodic growth and decay of the Be star disk, as suggested by \citep{we_lsi_xray}, based on the X-ray variability pattern. In this model, the shift of the maximum of the orbital X-ray light curve has been attributed to the confinement of the pulsar wind nebula by the Be star disk. The gradual growth of the disk leads to longer confinement of synchrotron-emitting electrons in the nebula and more pronounced synchrotron emission maximum at a later orbital phase, just before the nebula is deconfined when the pulsar exits from the disk. In this model, the system operates in two different modes. During a certain fraction of the orbit, the pulsar wind nebula is confined inside the disk and all emission coming from the system is from a rather compact region around the pulsar position. The second mode is when the pulsar wind nebula is deconfined and relativistic electrons can escape from the system, presumably along a bow-shaped contact surface between the pulsar and stellar winds, as in the model of Ref. \cite{1997ApJ...477..439T}.

The change of the variability pattern within a relatively narrow range between 0.1 and 1~GeV is surprising. The gamma rays at these two energies are most probably produced by the same mechanism, as indicated by the absence of any pronounced  break in the spectral energy distribution. It is thus not clear what effect might "erase" the orbital periodicity with the increase of the \gr\ energy. 

The \gr\ emission comes from an extended source that may have different spatial components, say the head and tail of the compact pulsar wind nebula. Particle acceleration conditions in these different components may be slightly different, so that the fractional contribution of these components to the overall \gr\ flux would be energy dependent. It  is possible that the change in the periodicity is explained by the different relative contributions of the different emission regions at different energies. For example, the head of the compact pulsar wind nebula may have a maximum emission in the periastron and provide a dominant contribution to the 100~MeV flux. The properties of the tail of the nebula may have a flux that depends on the characteristics  of the Be star disk. In this case, the orbital phase of the maximum flux from the tail may change in function of the disk size and orientation. If the tail component provides a sizeable contribution to the flux above 1~GeV, it would explain the appearance of the variability with the beat period in this energy range. 


\section{Conclusions}
\label{sec:conclusions}
In this paper, we report the energy dependence of $\gamma$-ray variability of \lsi.  We have shown that two different periods, $P_1 = 26.932\pm  0.004 (stat)\pm 0.008 (syst)$~d and $P_2 = 26.485 \pm 0.004 (stat)\pm 0.007 (syst)$~d, are detected in two energy ranges,  
$E>1$~GeV and  $E<0.3$~GeV. Within $1\sigma$ (stat+syst) the periods are consistent with orbital ($P_2$) and beat orbital/superorbital ($P_1$) periods. 

We have discussed the possible origin of the observed change in the periodicity over a factor of ten change in the energy. The presence of the maximum of the $0.1-0.3$~GeV orbit-folded light curve at the periastron can be explained, if this emission is produced via synchrotron or IC mechanisms. 

The re-appearance in $1-10$~GeV light curve of a new maximum at a phase that shifts cyclically on the superorbital timescale may point to the precession of one of the system components, for example, of the equatorial disk of the Be star. Emission from a part of the compact pulsar wind nebula that  provides a sizeable contribution to the GeV band flux may be affected by this precession. Alternatively, such variability can be connected to the process of a gradual build-up and decay of the Be star's disk on the superorbital time scale.
\subsubsection*{Acknowledgements}
 DM is supported by DFG through the grant MA 7807/2-1 and DLR through the grant 50OR2104. The authors acknowledge support by the state of Baden-W\"urttemberg through~bwHPC. The research conducted in this publication was jointly funded by the Irish Research Council under the IRC Ulysses Scheme 2021 and ministères français de l’Europe et des affaires étrangères (MEAE) et de l'enseignement supérieur et de la recherche (MESR).

\subsubsection*{Data Availability}
The data underlying this article will be shared on reasonable request to the corresponding authors.

\def\aj{AJ}%
\def\actaa{Acta Astron.}%
\def\araa{ARA\&A}%
\def\apj{ApJ}%
\def\apjl{ApJ}%
\def\apjs{ApJS}%
\def\ao{Appl.~Opt.}%
\def\apss{Ap\&SS}%
\def\aap{A\&A}%
\def\aapr{A\&A~Rev.}%
\def\aaps{A\&AS}%
\def\azh{AZh}%
\def\baas{BAAS}%
\def\bac{Bull. astr. Inst. Czechosl.}%
\def\caa{Chinese Astron. Astrophys.}%
\def\cjaa{Chinese J. Astron. Astrophys.}%
\def\icarus{Icarus}%
\def\jcap{J. Cosmology Astropart. Phys.}%
\def\jrasc{JRASC}%
\def\mnras{MNRAS}%
\def\memras{MmRAS}%
\def\na{New A}%
\def\nar{New A Rev.}%
\def\pasa{PASA}%
\def\pra{Phys.~Rev.~A}%
\def\prb{Phys.~Rev.~B}%
\def\prc{Phys.~Rev.~C}%
\def\prd{Phys.~Rev.~D}%
\def\pre{Phys.~Rev.~E}%
\def\prl{Phys.~Rev.~Lett.}%
\def\pasp{PASP}%
\def\pasj{PASJ}%
\def\qjras{QJRAS}%
\def\rmxaa{Rev. Mexicana Astron. Astrofis.}%
\def\skytel{S\&T}%
\def\solphys{Sol.~Phys.}%
\def\sovast{Soviet~Ast.}%
\def\ssr{Space~Sci.~Rev.}%
\def\zap{ZAp}%
\def\nat{Nature}%
\def\iaucirc{IAU~Circ.}%
\def\aplett{Astrophys.~Lett.}%
\def\apspr{Astrophys.~Space~Phys.~Res.}%
\def\bain{Bull.~Astron.~Inst.~Netherlands}%
\def\fcp{Fund.~Cosmic~Phys.}%
\def\gca{Geochim.~Cosmochim.~Acta}%
\def\grl{Geophys.~Res.~Lett.}%
\def\jcp{J.~Chem.~Phys.}%
\def\jgr{J.~Geophys.~Res.}%
\def\jqsrt{J.~Quant.~Spec.~Radiat.~Transf.}%
\def\memsai{Mem.~Soc.~Astron.~Italiana}%
\def\nphysa{Nucl.~Phys.~A}%
\def\physrep{Phys.~Rep.}%
\def\physscr{Phys.~Scr}%
\def\planss{Planet.~Space~Sci.}%
\def\procspie{Proc.~SPIE}%
\let\astap=\aap
\let\apjlett=\apjl
\let\apjsupp=\apjs
\let\applopt=\ao
\bibliography{biblio}

\end{document}